\documentstyle[aps,preprint,epsf,epsfig]{revtex}
\tightenlines
\begin{document}
\draft
\title{Teleportation Cost and 
Hybrid Compression of Quantum Signals}
\author{Masato Koashi and Nobuyuki Imoto}
\address{CREST Research Team for Interacting Carrier Electronics,
School of
Advanced Sciences, \\
The Graduate University for Advanced Studies (SOKEN),
Hayama, Kanagawa, 240-0193, Japan}
%\date{Mar 6, 2001}
\maketitle
\begin{abstract}
The amount of entanglement necessary to teleport quantum states 
drawn from general ensemble 
$\{p_i,\rho_i\}$ is derived. 
The case of perfect transmission of individual states and that of 
asymptotically faithful transmission are discussed. 
Using the latter result, we also derive 
the optimum compression rate when the ensemble is 
compressed into qubits and bits.

\end{abstract}
 
\narrowtext

The scheme of quantum teleportation~\cite{tele93}
realizes
transmission of quantum states between 
two remote parties connected only by a 
classical communication channel, in exchange for 
the consumption of a shared resource of entanglement.
This cost of entanglement is dependent on 
the sender's knowledge about the 
transferred quantum states. The two extreme cases 
are well known, namely,
 perfect 
transmission of one qubit in a completely unknown state
requires one unit (ebit) of entanglement,
and no entanglement is needed in the cases
where the sender 
can surely determine the state of the qubit.
Investigation for the cases in-between, where the sender is left with 
several possible states $\rho_i$ with probability 
$p_i$, has recently been started, and it was found that 
in the simplest case 
where $\{\rho_i\}$ consists of two nonorthogonal pure states,
perfect transmission requires one full ebit\cite{henderson00pra}.
What we derive in this Letter is the formula that gives  
the teleportation costs for arbitrary ensemble $\{p_i,\rho_i\}$,
where $\rho_i$ can be mixed states in a Hilbert space of any
finite size. The costs are obtained 
not only for the perfect transmission of individual samples
but also for the asymptotic case where independently drawn
samples are collectively teleported. 
This asymptotic result has a significant role in the field of 
quantum information theory, because it quantifies the amount 
of `purely quantum' or `nonclassical' information in the 
general quantum signals in the following sense.
A natural way to quantify the information in
a classical signal is
the least number of bits (Shannon entropy)
required to compress the signal reversibly~\cite{shannon48}.
Similarly, the information in a quantum signal 
can be measured by the least number of qubits required 
to be compressed~\cite{schumacher95,compress}. 
The present result pushes such argument one step further,
namely, it reveals how much portion of the qubits can be
replaced by classical bits.

\begin{figure}
\centerline {\epsfig{width=12.0cm,file=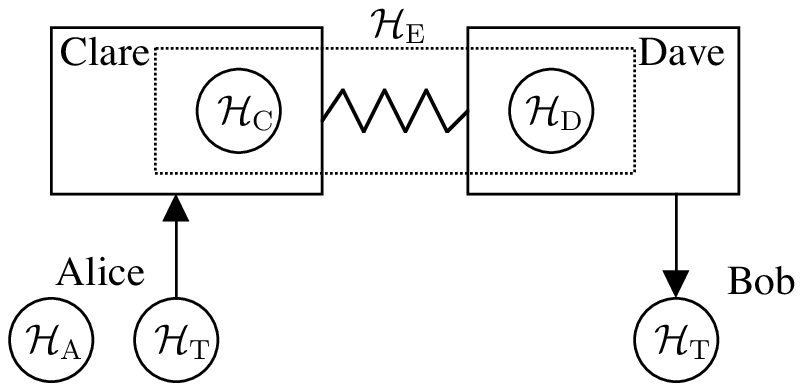}}
\caption{Roles of the four parties in the teleportation.
\label{fig1}}
\end{figure}

Since the problems considered here are sensitive 
to the information available in the teleportation process,
we separate the parties who conduct teleportation and 
those who prepare and receive quantum states.
We thus introduce four parties, namely, Alice prepares a state and 
give it to Clare, she teleports it to Dave, and he
finally delivers it to the receiver, Bob (see Fig.~\ref{fig1}). 
We first consider the {\em perfect} teleportation 
 of an ensemble ${\cal E}=\{p_i,\rho_i\}$. 
Suppose that Alice prepares the ensemble ${\cal E}=\{p_i,\rho_i\}$,
namely, she prepares a quantum system ${\cal H}_{\rm T}$
 (with a finite dimension
$d$) in a quantum state $\rho_i$ with (nonzero) probability $p_i$. 
She gives ${\cal H}_{\rm T}$ to Clare. Clare and Dave are separated from
each other and only the classical communication is allowed between them,
but they are supplied with entangled auxiliary systems 
${\cal H}_{\rm C}\otimes {\cal H}_{\rm D}$, where 
Clare owns ${\cal H}_{\rm C}$ and Dave has ${\cal H}_{\rm D}$.
Using this resource, the Clare-Dave team transfers the state 
$\rho_i$ and deliver the system ${\cal H}_{\rm T}$ to Bob. 
In the perfect teleportation, we require that `the received state of 
${\cal H}_{\rm T}$ for Bob is exactly $\rho_i$.' The precise
meaning of this statement is as follows. 
Suppose that the whole process was repeated many times,
 and after that, Alice disclosed the identity of each state 
(the number $i$) to Bob. He can do 
any statistical tests to check whether the state $\rho_i$ was 
correctly transferred, and what we require is that 
the delivered states must 
pass any such tests.

If we take notice of the input and the output of 
the task done by the Clare-Dave team, their task can be
regarded as 
a quantum operation on ${\cal H}_{\rm T}$. This operation is
independent of Alice's choice of the number $i$, and
maps any $\rho_i$ to $\rho_i$. We have a useful theory \cite{what} 
for such operations. 
To state the results of this theory, 
it is convenient to express quantum operations
in unitary representation, namely, by unitary operations $U$ 
acting on the combined space ${\cal H}_{\rm T}\otimes {\cal H}_{\rm E}$, 
where ${\cal H}_{\rm E}$ represents an auxiliary system initially 
prepared in a standard pure state $\Sigma_{\rm E}$. 
In the present context, ${\cal H}_{\rm E}$ includes 
${\cal H}_{\rm C}\otimes {\cal H}_{\rm D}$, the communication
line, and any other auxiliary systems.
Then, 
it was shown \cite{what} that,
given $\{\rho_i\}$,
we can find a decomposition of 
${\cal H}_{\rm Tsup}$ defined as the support of $\sum_i\rho_i$
(${\cal H}_{\rm Tsup}$ is generally a subspace of 
${\cal H}_{\rm T}$) written as
\begin{equation}
{\cal H}_{\rm Tsup}=\bigoplus_l
{\cal H}^{(l)}_{\rm J} \otimes {\cal H}^{(l)}_{\rm K}, 
\label{hdec}
\end{equation}
in such a way that any quantum operation preserving $\{\rho_i\}$
is expressed in the following form
\begin{equation}
U({\bf 1}_{\rm Tsup}\otimes \Sigma_{\rm E})
=\bigoplus_l\bbox{1}^{(l)}_{\rm J}\otimes 
U^{(l)}_{\rm KE}({\bf 1}^{(l)}_{\rm K}\otimes \Sigma_{\rm E}), 
\label{main1}
\end{equation}
where $U^{(l)}_{\rm KE}$ are unitary operators acting on the combined
space ${\cal H}^{(l)}_{\rm K}\otimes {\cal H}_{\rm E}$.
Under this decomposition, $\rho_i$
is written as
\begin{equation}
\rho_i=\bigoplus_l q^{(i,l)} \rho^{(i,l)}_{\rm J}\otimes \rho^{(l)}_{\rm
K},
\label{rhodec}
\end{equation}
where $\rho^{(i,l)}_{\rm J}$ and $\rho^{(l)}_{\rm K}$ are normalized density 
operators acting on ${\cal H}^{(l)}_{\rm J}$ and ${\cal H}^{(l)}_{\rm K}$, respectively, 
and
$q^{(i,l)}$ is the probability for the state to be  in the subspace
${\cal H}^{(l)}_{\rm J}\otimes{\cal H}^{(l)}_{\rm K}$.
 $\rho^{(l)}_{\rm K}$ is independent  of $i$, and
$\{\rho^{(1,l)}_{\rm J},\rho^{(2,l)}_{\rm J},\ldots\}$ cannot be
expressed in a simultaneously block-diagonalized form. 
An explicit procedure to obtain this decomposition is also 
given in \cite{what}.
The total density operator $\rho\equiv \sum_i p_i \rho_i$ for the
ensemble ${\cal E}$ is also decomposed as 
\begin{equation}
\rho=\bigoplus_l p^{(l)} \rho^{(l)}_{\rm J}\otimes \rho^{(l)}_{\rm K},
\label{totalrhodec}
\end{equation}
where $p^{(l)}\equiv\sum_i p_i q^{(i,l)}$ and 
$\rho^{(l)}_{\rm J}\equiv (\sum_i p_i q^{(i,l)}
\rho^{(i,l)}_{\rm J})/p^{(l)}$. The von Neumann entropy
of $\rho$, defined as $S(\rho)\equiv -\mbox{Tr}\rho \log_2 \rho$,
 is written by the sum of three parts as follows,
\begin{eqnarray}
S(\rho)&=&
\sum_l p^{(l)}\left[
-\log_2 p^{(l)}+S(\rho^{(l)}_{\rm J})+
S(\rho^{(l)}_{\rm K})
\right]
\nonumber \\
&\equiv& I_{\rm C}+I_{\rm NC}+I_{\rm R},
\label{vondec}
\end{eqnarray}
where each part is uniquely determined as a function of ${\cal E}$.

The decomposition (\ref{rhodec}) immediately gives us one  
scheme of perfect teleportation: Clare measures the value 
$l$ and tells the result to Dave. Then she teleports  
$\rho^{(i,l)}_{\rm J}$ to Dave. Knowing the value $l$ and 
having the state $\rho^{(i,l)}_{\rm J}$, Dave can locally 
prepare ${\cal H}_{\rm T}$ in the state
$\rho^{(i,l)}_{\rm J}\otimes \rho^{(l)}_{\rm K}$.
Dave finally delivers this state to Bob without disclosing the
 identity of $l$, resulting in the delivery of the state $\rho_i$
to Bob.
 In this scheme, the amount of entanglement
$E_{\rm per}^{(p)}$
measured in units of ebits (number of Bell pairs) that must 
be {\em prepared} in ${\cal H}_{\rm C}\otimes {\cal H}_{\rm D}$
beforehand is 
\begin{equation}
E_{\rm per}^{(p)}=\log_2\max_l{\rm dim}\;{\cal H}^{(l)}_{\rm J}.
\label{Eperpre}
\end{equation}
In fact, we can show that $E_{\rm per}^{(p)}$ is the lower bound for {\em
any}  perfect teleportation scheme. Suppose that Alice has her 
auxiliary system ${\cal H}_{\rm A}=\bigoplus_l {\cal H}_{\rm A}^{(l)}$
with ${\rm dim} {\cal H}^{(l)}_{\rm A}={\rm dim} {\cal H}^{(l)}_{\rm J}$,
and let $|\Phi_{\rm max}^{(l)}\rangle$ be a maximally entangled 
state in the subspace ${\cal H}^{(l)}_{\rm A}\otimes{\cal H}^{(l)}_{\rm
J}$. Instead of preparing the state drawn from ${\cal E}$, 
Alice can prepare
$|\Phi_{\rm max}^{(l)}\rangle$ with the choice of $l$ that gives the
maximum in Eq.~(\ref{Eperpre}), and  send ${\cal H}_{\rm T}$ to Clare.
The form (\ref{main1}) implies that any perfect teleportation scheme for
$\{\rho_i\}$ teleports any state in ${\cal H}^{(l)}_{\rm J}$
correctly. Hence Alice and Bob can share the state 
$|\Phi_{\rm max}^{(l)}\rangle$ without fail, which amounts 
to the entanglement of $E_{\rm per}^{(p)}$. Noting
that  the Alice-Clare group and the Dave-Bob group are only connected
by the classical communication line, we conclude  $E_{\rm
per}^{(p)}$ ebits of entanglement is the minimum that must be prepared in 
${\cal H}_{\rm C}\otimes {\cal H}_{\rm D}$ beforehand.

While $E_{\rm per}^{(p)}$ ebits must be prepared to assure the 
perfect teleportation, the whole entanglement does not necessarily
consumed. In the particular scheme explained above, the consumption
obviously depends on the measurement result of $l$, which follows 
the probability distribution $p^{(l)}$.
Considering the situation where the teleportation for ${\cal E}$
is repeated many times, we can define
the average consumption of entanglement per system.
For the above scheme, this is given by 
\begin{equation}
E_{\rm per}^{(c)}=\sum_l p^{(l)}\log_2{\rm dim}\;{\cal H}^{(l)}_{\rm J}.
\end{equation}
This value again can be shown to be the lower bound for any 
perfect teleportation scheme,
by considering the case in which Alice prepares the state
$\bigoplus_l p^{(l)}|\Phi_{\rm max}^{(l)}\rangle\langle\Phi_{\rm
max}^{(l)}|
\otimes \rho^{(l)}_{\rm K}$.
The form (\ref{main1}) ensures that 
after the teleportation, $E_{\rm per}^{(c)}$ ebits
of entanglement are shared between Alice and Bob, and that 
the state left in 
${\cal H}_{\rm E}$  is identical to the case in which Alice uses 
${\cal E}$. Therefore, at least $E_{\rm per}^{(c)}$ ebits
on average must be consumed in the perfect teleportation of ${\cal E}$.
 
Next, we consider the asymptotically faithful teleportation of ${\cal E}$.
In contrast to the individual teleportation described above, 
we allow collective manipulations of the sequence of messages. 
In this case, Alice prepares $N$ messages drawn from ${\cal E}$,
namely, she prepares the state 
$\rho^N_\lambda\equiv \rho_{i_1}\otimes\cdots\otimes\rho_{i_N}$
acting on a Hilbert space 
${\cal H}_{\rm T}^N\equiv {\cal H}_{{\rm T}1}\otimes\cdots\otimes{\cal
H}_{{\rm T}N}$ with probability $p^N_\lambda= p_{i_1}\ldots p_{i_N}$,
where
$\lambda$ represents a set of indexes $\{i_1,\ldots ,i_N\}$.
Clare receives ${\cal H}_{\rm T}^N$ from Alice and teleports the required
information to Dave, who delivers to Bob the state 
$\rho^\prime_\lambda$ acting on ${\cal H}_{\rm T}^N$.
The quality of the scheme can be measured by 
the average fidelity 
\begin{equation}
\bar{F}\equiv\sum_\lambda 
p^{N}_\lambda F(\rho^N_\lambda, \rho^\prime_\lambda),
\end{equation}
where the fidelity function \cite{jozsa94fid} is defined as
$F(\rho,\sigma)\equiv [\mbox{Tr}\sqrt{\rho^{1/2}\sigma\rho^{1/2}}]^2$.
The teleportation cost for ${\cal E}$ in this case can be 
defined as 
the bound $E_{\rm asy}$ for the amount of entanglement per message
such that for arbitrary 
small $\delta>0$, (a) if $E_{\rm asy}+\delta$ 
ebits per message is given,
we can find a sequence of schemes with 
$\bar{F}\rightarrow 1 (N\rightarrow \infty)$,
and (b) if $E_{\rm asy}-\delta$ ebits per message is
given, no such sequences exist.

The decomposition (\ref{rhodec}) again helps to 
build one example of  asymptotically faithful teleportation,
which is stated as follows.
Clare measures the value 
$l$ for each message, and tells Dave the result
$\{l_1,\ldots ,l_N\}$.
Then they classify the messages according to the value of $l$.
Each group
contains  
$p^{(l)}N$ messages on average. By ignoring ${\cal H}^{(l)}_{\rm K}$,
the postmeasurement state of this group of messages can be regarded as 
the one drawn from the ensemble $\{
 p_i q^{(i,l)}/p^{(l)},\rho^{(i,l)}_{\rm J}\}$ (with fixed $l$).
Clare compresses each group of messages (classified by $l$)
into $S(\rho^{(l)}_{\rm J})$ qubits per message\cite{schumacher95,lo95}, 
and teleports it to Dave. Dave decompresses it and attaches
$\rho^{(l)}_{\rm K}$ locally, and delivers to Bob after ordering 
the messages correctly using the information $\{l_1,\ldots ,l_N\}$.
This strategy becomes asymptotically faithful if $I_{\rm NC}+\delta$
Bell pairs per message are given.
The existence of this example assures $E_{\rm asy}\le I_{\rm NC}$.

We can also obtain the opposite inequality $E_{\rm asy}\ge I_{\rm NC}$ 
in the following way. 
The form of Eq.~(\ref{rhodec}) implies that the spaces ${\cal
H}^{(l)}_{\rm K}$ are redundant in the ensemble ${\cal
E}=\{p_i,\rho_i\}$.  Consider the ensemble  
${\cal E}_{\rm R}\equiv\{p_i,\bigoplus_l
q^{(i,l)}
\rho^{(i,l)}_{\rm J}\}$, in which the redundancy has been removed.
The two ensembles ${\cal E}$ and ${\cal E}_{\rm R}$
are completely interchangeable, namely, there exist quantum operations
that converts one to the other. Then, we can easily prove that 
$E_{\rm asy}({\cal E})=E_{\rm asy}({\cal E}_{\rm R})$
\cite{horodecki98,compress}. Since 
$I_{\rm NC}({\cal E})=I_{\rm NC}({\cal E}_{\rm R})$,
it is suffice to consider the cases where $\{\rho_i\}$
have no redundancy, namely, ${\cal E}_{\rm R}={\cal E}$,
in proving $E_{\rm asy}\ge I_{\rm NC}$. One benefit of
this assumption is that the 
decomposition (\ref{hdec}) and the requirement (\ref{main1}) 
can be simplified as ${\cal H}_{\rm Tsup}=
\bigoplus_l{\cal H}^{(l)}_{\rm J}$  and 
\begin{equation}
U({\bf 1}_{\rm Tsup}\otimes \Sigma_{\rm E})
=\bigoplus_l\bbox{1}^{(l)}_{\rm J}\otimes 
U^{(l)}_{\rm E}\Sigma_{\rm E}, 
\label{unec}
\end{equation}
where $U^{(l)}_{\rm E}$ are unitary operators acting on 
${\cal H}_{\rm E}$. 
Let us introduce a measure $f(U)$, which characterizes
how well an arbitrary $U\in U(d+d^2)$ acting on 
${\cal H}_{\rm T}\otimes {\cal H}_{\rm E}$ preserves
the states in ${\cal H}_{\rm T}$ drawn from ${\cal E}$.
It is defined as the nonnegative function
$f(U)\equiv 1-\sum_i p_i F(\rho_i,\Lambda_U (\rho_i))$,
where $\Lambda_U (\rho_i)\equiv {\rm Tr_E}[U(\rho_i\otimes 
\Sigma_{\rm E})U^\dagger]$.
Since $f(U)=0$ iff  $\Lambda_U (\rho_i)=\rho_i$ for all $i$,
$f^{-1}(0)$ is equal to the set of $U$ that can be expressed in 
the form (\ref{unec}).

In the teleportation of $N$ messages,
the whole operation done by Clare and Dave
should be  written as a quantum operation $\rho^N_\lambda\rightarrow
\rho^\prime_\lambda =\Lambda(\rho^N_\lambda)$.
In this process, the marginal state in the first system 
(${\cal H}_{{\rm T}1}$)
evolves from $\rho_{i_1}$ to $\mbox{Tr}_{2\ldots N}(\rho^\prime_\lambda)$.
This evolution can be regarded as a result of a quantum operation 
$\Lambda_1$, defined as 
\begin{eqnarray}
\Lambda_1 (\rho_i)&\equiv& \sum_{i_2\ldots i_N}
 p_{i_2}\ldots p_{i_N}\mbox{Tr}_{2\ldots N}
\Lambda (\rho_i\otimes\rho_{i_2}\otimes\cdots\otimes\rho_{i_N})
\nonumber \\
&=&\mbox{Tr}_{2\ldots N}
\Lambda (\rho_i\otimes\rho\otimes\cdots\otimes\rho).
\label{deflambda1}
\end{eqnarray}
Note that $\Lambda_1$ is determined by $\Lambda$ and the {\em total}
density 
operators ($\rho$) of the initial state ensembles of 
the other $N-1$ systems. Let us take a unitary representation 
of $\Lambda_1$ as $U_1\in U(d+d^2)$ acting on 
${\cal H}_{{\rm T}1}$ and an auxiliary system ${\cal H}_{{\rm E}1}$.
Here we can assume the dimension of ${\cal H}_{{\rm E}1}$ to be 
$d^2$ \cite{schumacher96q}.
From the properties of the fidelity function $F$, we obtain 
\begin{eqnarray}
\bar{F}&=&\sum_\lambda 
p^{N}_\lambda F(\rho^N_\lambda, \rho^\prime_\lambda)
\nonumber \\
&\le&
\sum_\lambda p^N_\lambda F(\rho_{i_1},\mbox{Tr}_{2\ldots N}
\Lambda (\rho_{i_1}\otimes\rho_{i_2}\otimes\cdots\otimes\rho_{i_N}))
\nonumber \\
&\le&\sum_i p_i F(\rho_i,\Lambda_1 (\rho_i))=1-f(U_1)
\label{fidbound}
\end{eqnarray}
since the fidelity does not decrease under partial trace
(the first inequality) and $F(\sigma,\rho)$ is convex 
as a function of $\rho$ (the second). 

The next step is to consider what happens if, 
instead of drawing from the ensemble ${\cal E}$, Alice prepares 
an entangled states $\rho_{\rm AT}\equiv\bigoplus_l p^{(l)} 
|\Psi^{(l)}\rangle\langle \Psi^{(l)}|$ in 
${\cal H}_{\rm A}\otimes{\cal H}_{\rm T}$.
Here we assume that $|\Psi^{(l)}\rangle$ is 
a pure state in the subspace
${\cal H}^{(l)}_{\rm A}\otimes{\cal H}^{(l)}_{\rm J}$
and satisfies ${\rm Tr}_{\rm A}
(|\Psi^{(l)}\rangle\langle \Psi^{(l)}|)=\rho^{(l)}_{\rm
J}$.  
Note that $S(\rho_{\rm AT})=I_{\rm C}$ and 
$S({\rm Tr}_{\rm T}[\rho_{\rm AT}])=I_{\rm C}+I_{\rm NC}$.
Suppose that Alice prepares $N$ identical states 
$\rho_{\rm AT}$ in ${\cal H}_{{\rm A}k}\otimes{\cal H}_{{\rm T}k}
(k=1,\ldots,N)$ and gives the system ${\cal H}_{\rm T}^N$
to Clare, who conducts the operation $\Lambda$ with Dave.
Since the marginal state in ${\cal H}_{{\rm T}k}
(k=2,\ldots,N)$  is $\rho$,
which is identical to the state from ${\cal E}$, the operation 
on ${\cal H}_{{\rm T}1}$ is again given by $\Lambda_1$. 
Under this operation, the total state in  
${\cal H}_{{\rm A}1}\otimes{\cal H}_{{\rm T}1}$ evolves from 
$\rho_{\rm AT}$ to 
$\tilde\rho_{\rm AT}={\rm Tr}_{\rm E}[({\bf 1}_{{\rm A}1}
\otimes U_1)(\rho_{\rm AT}\otimes \Sigma_{\rm E})
({\bf 1}_{{\rm A}1}
\otimes U_1^\dagger)]$. 
The entropy production in this process can be 
regarded as a function of $U_1$, namely,
$g(U_1)\equiv S(\tilde\rho_{\rm AT})-S(\rho_{\rm AT})$.
This function is related to the function $f$ as follows. 
Since the form (\ref{unec}) preserves $\rho_{\rm AT}$, 
$g(U_1)$ is zero for any $U_1 \in f^{-1}(0)$.
Let us define the set 
$\bar{X}_\delta\equiv\{U|g(U)\ge \delta\}$ for arbitrary $\delta>0$.
Since $g$ is continuous, $\bar{X}_\delta$ is a closed subset of 
$U(d+d^2)$.
Since $U(d+d^2)$ is compact and $f$ is continuous, 
the image
$f(\bar{X}_\delta)$ is closed in $R$. 
$\bar{X}_\delta \cap f^{-1}(0)=\emptyset$ 
implies that $0\notin f(\bar{X}_\delta)$.
Therefore, $f(\bar{X}_\delta)$ has its minimum $\eta(\delta)>0$.
Note that the functional dependence of $\eta$ on $\delta$ is 
determined by ${\cal E}$, and is independent of $N$.
This result will be used below.

After the teleportation of $N$ systems, Alice and Bob 
obtain an entangled state. The entanglement of formation,
$E_f$, satisfies the inequality $E_f\ge S_{\rm A}-S_{\rm AB}$,
where $S_{\rm A}$ and $S_{\rm AB}$ are the von Neumann entropy 
of Alice's marginal state and that of the whole state
of the $N$ systems, respectively.
(This inequality is shown in \cite{henderson00prl}
without details of the proof. Alternatively, it can be proved by
applying the strong subadditivity~\cite{Lieb73}
of $S$ to the state 
$\rho^\epsilon_{\rm ABM}$ defined in \cite{henderson00prl}.)
Since Alice's marginal state does not change in the teleportation,
$S_{\rm A}$ is given by $S_{\rm A}=N(I_{\rm C}+I_{\rm NC})$. From the
subadditivity of $S$, we have $S_{\rm AB} \le \sum_k [g(U_k) +
S(\rho_{\rm AT})]\le Ng(U_1) + NI_{\rm C}$,
where we have assumed that 
$g(U_1)\ge g(U_k)$ for all $k$ 
without loss of generality,
since the numbering of the
systems 
${\cal H}_{{\rm T}k}$ is arbitrary.
Combining these, we obtain 
$E_f/N\ge I_{\rm NC}-g(U_1)$. Now, suppose that 
Clare and Dave used $I_{\rm NC}-\delta$
Bell pairs per message. Since the entanglement of formation 
of the total system never increases, $I_{\rm NC}-\delta\ge E_f/N$.
This gives $g(U_1)\ge \delta$. As shown above, this means  
$f(U_1)\ge \eta(\delta)$. Combined with (\ref{fidbound}),
we find that the fidelity is bounded from above as 
$\bar{F}\le 1-\eta(\delta)<1$, regardless of $N$. 
This implies that faithful teleportation is impossible 
with this resource of Bell pairs.
We have thus proved the inequality $E_{\rm asy}\ge I_{\rm NC}$.
As explained before, this inequality also holds in general cases
where ${\cal E}\neq {\cal E}_{\rm R}$.
Combined with the opposite inequality, 
we conclude that the cost 
for asymptotically faithful teleportation 
is given by $E_{\rm asy}= I_{\rm NC}$.

In the example of asymptotically faithful teleportation scheme
mentioned above, Dave reproduces ${\cal E}$ from 
$I_{\rm NC}$ qubits per message received by teleportation 
and the classical information of 
the measurement result $\{l_1,\ldots ,l_N\}$, which can be 
compressed into $I_{\rm C}$ bits per message\cite{shannon48}. 
This implies that for arbitrary small $\delta_1>0$ and 
$\delta_2>0$, the quantum signal ${\cal E}$ can be
 compressed and decompressed back faithfully in the 
asymptotic limit, if $I_{\rm NC}+\delta_1$ qubits and 
$I_{\rm C}-\delta_1+\delta_2$ bits per message are available
(note that a qubit can substitute for a bit).
This is indeed the optimal way of compression into 
bits and qubits in the blind protocols, in the following 
sense: (a) The result for the teleportation cost
 derived 
above, $E_{\rm asy}= I_{\rm NC}$, 
 implies that if only $I_{\rm NC}-\delta$ qubits per message
(and any number of bits) are available, no faithful compression is
possible. (b) $I_{\rm C}+I_{\rm NC}=S(\bigoplus_l p^{(l)}\rho^{(l)}_{\rm
J})$  is equal to the passive information 
$I_{\rm p}$ \cite{compress}. 
This means that 
if $I_{\rm p}-\delta$ qubits (and no bits) are available,
no faithful compression is possible. It is thus 
impossible to conduct faithful compression 
if the available number of bits plus that of qubits per message
is $I_{\rm C}+I_{\rm NC}-\delta$.

In summary, for general ensembles of quantum states 
including mixed states, we have derived 
the amount of entanglement required to teleport them 
and the optimum number of bits and qubits necessary to 
compress them faithfully. Together with the earlier 
results \cite{what} of no-cloning or no-imprinting 
argument, the present results elucidate the 
property of information stored in a quantum ensemble
as follows. The information can always be decomposed 
into three parts, namely, redundant, classical, and 
nonclassical parts. The classical part can be copied,
compressed into bits, and transmitted through a classical 
channel. The nonclassical part of the information cannot be
extracted without disturbing the states, 
can be compressed only into qubits, and must consume entanglement
when transmitted over a classical channel.

This work was supported by a Grant-in-Aid for Encouragement of Young
Scientists (Grant No.~12740243) and a Grant-in-Aid for Scientific 
Research (B) (Grant No.~12440111)
by the Japan Society of the Promotion of
Science.

\newpage

\end{document}